\documentclass[a4paper,11pt]{article}
\usepackage{pos}
\usepackage{float}

\title{NLO heavy-quark contributions to DIS structure functions in the ACOT scheme}

\author*[a]{E. Spezzano}
\author[a]{T.~Je\v{z}o}
\author[a]{M.~Klasen}
\author[b,c]{P.~Risse}
\author[d]{I.~Schienbein}

\usepackage[firstpage]{draftwatermark}
\SetWatermarkFontSize{9pt}
\SetWatermarkColor{black}
\SetWatermarkAngle{0}
\SetWatermarkHorCenter{17.17cm}
\SetWatermarkVerCenter{5.8cm}
\SetWatermarkText{}
\SetWatermarkText{MS-TP-25-14, SMU-PHY-25-04, JLAB-THY-25-4606}

\affiliation[a]{Institut  für  Theoretische  Physik,  Universität  Münster,  Wilhelm-Klemm-Straße 9, 48149 Münster, Germany}
\affiliation[b]{Department of Physics, Southern Methodist University, Dallas, TX 75275-0175, U.S.A.}
\affiliation[c]{Theory Center, Jefferson Lab, Newport News, VA 23606, U.S.A.}
\affiliation[d]{Laboratoire de Physique Subatomique et de Cosmologie, Université Grenoble-Alpes,CNRS/IN2P3, 53 Avenue des Martyrs, 38026 Grenoble, France}

\emailAdd{edoardo.spezzano@uni-muenster.de}

\abstract{
We present next-to-leading-order (NLO) calculations of heavy-quark contributions to deep-inelastic scattering (DIS) structure functions $F_4$ and $F_5$ within the Aivazis--Collins--Olness--Tung (ACOT) scheme, implemented in the open source library \texttt{APFEL++} using \texttt{CT18NLO}  parton distribution functions. These structure functions, suppressed by lepton mass effects in light-lepton processes, become significant in muon, tau-lepton and neutrino scattering at facilities such as SHiP, IceCube, and DUNE. Our results reveal NLO corrections up to 10\% relative to leading order, with pronounced heavy-quark effects at low Bjorken-$x$, impacting gluon and strange quark distributions. In the unpolarized case, $F_{4/5}^{\gamma Z}$ and $F_{4/5}^{\gamma}$ do not contribute to the cross section, while the $\gamma Z$ interference becomes accessible with longitudinally polarized lepton beams at the Electron-Ion Collider (EIC), offering enhanced sensitivity at low $Q^2$ due to reduced $Z$-boson propagator suppression. Analytical NLO expressions have also been derived for the polarized structure functions $g_1$, $g_4$, $g_5$, $g_6$, and $g_7$ in the ACOT framework. These developments enable precise theoretical predictions for upcoming experimental programs and global QCD analyses.
}

\FullConference{XXXII International Workshop on Deep Inelastic Scattering and Related Subjects (DIS2025)\\
24-28 March, 2025\\
Cape Town, South Africa\\}

\begin{document}
\tableofcontents
\maketitle

\section{Introduction}
Deep-inelastic scattering (DIS) is a cornerstone of quantum chromodynamics (QCD), providing precise probes of nucleon and nuclear structure through high-energy lepton or neutrino interactions. The structure functions $F_1$, $F_2$, and $F_3$ are well-studied, while $F_4$ and $F_5$ have never been measured due to suppression by kinematic factors proportional to squared lepton mass \citep{ParticleDataGroup:2024cfk}. These functions become significant in processes involving heavy leptons, such as muon and tau leptons, which will be accessible in the near future. Recent advancements, including the proposed Search for Hidden Particles (SHiP) experiment at CERN \citep{shipcollaboration2022shipexperimentproposedcern}, tau neutrino detections at IceCube \citep{IceCube:2024nhk}, the upcoming Deep Underground Neutrino Experiment (DUNE) \citep{abi2020deepundergroundneutrinoexperiment}, and polarized DIS experiments at the Electron-Ion Collider (EIC) \citep{khalek2022snowmass2021whitepaper}, highlight the need for precise theoretical predictions. In particular, SHiP, with beam energies yielding $Q^2$ up to 10--100\,GeV$^2$, will be able to test low $x$ values between 0.08 and 0.3 \citep{shipcollaboration2022shipexperimentproposedcern}, and DUNE, probing lower energy regions with $Q^2$ between 1 and 20\,GeV$^2$ and $x$ values between 0.01 and 0.50 \citep{abi2020deepundergroundneutrinoexperiment}, will test distinct kinematic regimes. Heavy quarks play a critical role in DIS, particularly at low Bjorken-$x$, offering insights into gluon and strange quark distributions \citep{Stavreva:2012bs}. We present next-to-leading-order (NLO) calculations of heavy-quark contributions to the unpolarized structure functions $F_4$ and $F_5$ using the Aivazis--Collins--Olness--Tung (ACOT) scheme \citep{Aivazis:1993kh,Aivazis:1993pi}, implemented in \texttt{APFEL++} \citep{Bertone:2016lga} which is a versatile numerical library for high-precision QCD calculations, offering efficient and accurate evaluation of structure functions. Analytical NLO calculations have also been completed for the longitudinally polarized structure functions $g_1$, $g_4$, $g_5$, $g_6$, and $g_7$ within the same ACOT framework, though numerical results are not yet available.

\section{Theoretical framework}
Heavy-quark production plays a central role in DIS, particularly for understanding gluon and strange quark dynamics at low $x$. The ACOT scheme integrates heavy-quark mass effects within a variable flavor number framework, ensuring consistency across kinematic regions from $Q \sim m_H$ and $Q \gg m_H$, where $m_H$ is the heavy quark mass \citep{Collins:1998rz}. We extend prior calculations for $F_1$, $F_2$, and $F_3$ \citep{Kretzer:1998ju} to include $F_4$ and $F_5$, providing a unified framework for unpolarized DIS essential for interpreting modern experimental data.

\section{Unpolarized structure functions}
We calculate NLO contributions to the unpolarized structure functions $F_4$, and $F_5$ within the ACOT scheme. The cross sections for neutral current (NC) and charged current (CC) processes are given by:
\begin{align}
\frac{d^2\sigma^{ZZ}}{dx\,dy} 
&= \frac{4\pi\alpha^2}{x y Q^2} \, \eta^{ZZ} \Bigg\{ \,
\frac{x y^2 \Big( (g_A^{e \, 2}+g_{V}^{e \, 2}) Q^2 + (6 g_A^{e \,2} - 2 g_{V}^{e \, 2})\mu_1^2 \Big)}{Q^2} \, F_1^{Z} \nonumber \\[6pt]
&\quad - \frac{ \Big( (g_{A}^{e \, 2}+g_{V}^{e \, 2})\big((y-1)Q^2 + M^2 x^2 y^2\big) Q^2 + 4 g_{A}^{e \, 2} M^2 x^2 y^2 \mu_1^2 \Big)}{Q^4} \, F_2^{Z} \nonumber \\[6pt]
&\quad -  g_A^e  g_V^e\, x \, y (y-2) \, F_3^{Z} 
+ \frac{2 g_A^{e\, 2} x y^2 \mu_1^2}{Q^2} \left( F_4^{Z} - \, F_5^{Z} \right)
\Bigg\}, 
\label{eqn: ZZ cross section}
\end{align}
and:
\begin{align}
\frac{d^2\sigma^{WW}}{dx\,dy} &= \frac{4\pi\alpha^2}{x y Q^2} \tilde{\eta}\eta^W \Bigg\{ \,
\frac{x y^2 \big(Q^2 + \mu_1^2 + \mu_2^2\big)}{Q^2} \, F_1^{W}  - \frac{2 \Big( (y-1) Q^4 + M^2 x^2 y^2 (Q^2 + \mu_1^2 + \mu_2^2) \Big)}{Q^4} \, F_2^{W} \nonumber \\[6pt] 
&\quad \mp\frac{x y \big( (2-y)Q^2 + y (\mu_1^2- \mu_2^2) \big)}{Q^2} \, F_3^{W}  + \frac{x y^2 \Big( \mu_1^4 + (Q^2 - 2 \mu_2^2)\mu_1^2 + \mu_2^2(Q^2 + \mu_2^2) \Big)}{Q^4} \, F_4^{W} \nonumber \\[6pt]
&\quad - \frac{2x y \big( (y-1)\mu_1^2 + \mu_2^2 \big)}{Q^2} \, F_5^{W}
\Bigg\},
\label{eqn: CC cross section}
\end{align}
where $\mu_1$ and $\mu_2$ are the masses of incoming and outgoing leptons, $\tilde{\eta}=2$ for (anti)neutrinos, and $\eta^Z$ and $\eta^W$ are the neutral- and charged-current propagators as detailed in Ref. \citep{ParticleDataGroup:2024cfk}. The couplings $g_A^e$ and $g_V^e$ represent the axial and vector couplings of the $Z$ boson to the incoming lepton, as detailed in \citep{ParticleDataGroup:2024cfk}. 

In the unpolarized case, the $\gamma \gamma$ and $\gamma Z$ interference terms don't contribute to the cross section for $F_4$ and $F_5$, making $F_{4/5}^{\gamma Z}$ and $F_{4/5}^{\gamma}$ irrelevant. However, with longitudinally polarized incoming leptons, as planned for the EIC \citep{khalek2022snowmass2021whitepaper}, the $\gamma Z$ interference term becomes accessible in the polarized cross section, while the pure photon contribution remains absent. This is crucial because the $\gamma Z$ interference is less suppressed by the $Z$ boson propagator at low $Q^2$ than the pure $ZZ$ term, potentially enhancing the sensitivity to the $F_{4/5}^{\gamma Z}$ in experiments with polarized beams. The same argument applies to the structure function $F_6$, which can, in principle, be accessed experimentally. In the Standard Model, however, $F_6$ is expected to vanish; therefore, any non-zero measurement would constitute evidence for CP-violating effects. Detecting such a deviation would require precise experimental determination and theoretical calculation of the cross section, further emphasizing the importance of accurately understanding and constraining $F_4$ and $F_5$. Thus, we focus on $F_{4/5}^{ZZ/WW}$, which dominate the unpolarized NC/CC structure functions. The equations in Eqs. \eqref{eqn: ZZ cross section} and \eqref{eqn: CC cross section} highlight the suppression of $F_4$ and $F_5$ due to lepton mass effects, making them measurable in heavy-lepton processes like those at SHiP or IceCube. Our NLO calculations show corrections up to 10\% of leading-order terms for $F_5$, with heavy quarks significantly influencing gluon and strange quark distributions at low $x$. 

\section{Numerical results}
Using \texttt{APFEL++} with \texttt{CT18NLO} PDFs \citep{Hou_2021}, we evaluate the unpolarized structure functions $F_4$ and $F_5$ for the charged current ($W^{\pm}$) and neutral current ($ZZ$) channels. These implementations will enable global analyses of structure functions and comparisons with experimental data from SHiP, IceCube, and DUNE. We report here the numerical results for $F_{4/5}^{\rm NC/CC}(x,Q^2)$ structure functions.

\begin{figure}[H]
    \centering
    \includegraphics[width=0.5\textwidth]{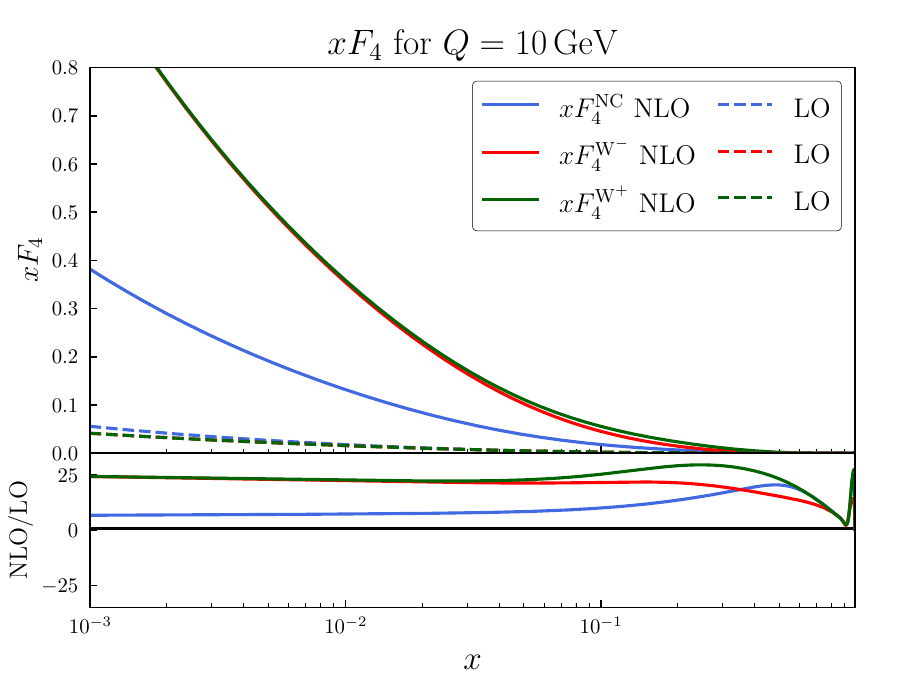}%
    \includegraphics[width=0.5\textwidth]{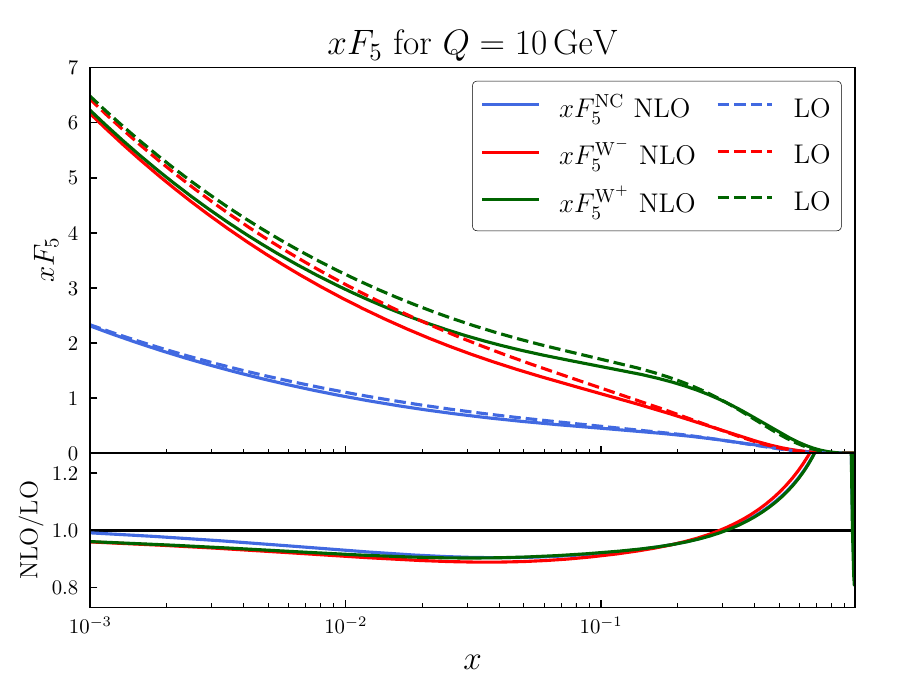}
 \caption{
Comparison of the structure functions $F_4$ (left) and $F_5$ (right) for $Q = 10\,\mathrm{GeV}$ 
as functions of the Bjorken variable $x$ in the range $10^{-3} \le x \le 1$. 
Solid lines correspond to next-to-leading order (NLO) results and dashed lines to leading order (LO), 
for both neutral-current and charged-current interactions. 
}

    \label{fig:F4F5}
\end{figure}

As shown in Fig.~\ref{fig:F4F5}, the NLO corrections to $F_4$ are significantly larger than the LO contribution. 
This behavior originates from the Albright-Jarlskog (AJ) relations \citep{ALBRIGHT1975467}, which imply that $F_4$ vanishes at leading order in the massless quark limit. In our calculation, however, the LO prediction for $F_4$ is not exactly zero because the finite masses of the heavy quarks are taken into account. In contrast, $F_5$ receives a LO contribution, and the NLO corrections are comparatively moderate. 
Nevertheless, deviations between the LO and NLO predictions for $F_5$ can reach up to about $10\%$, indicating that higher-order effects still play a non-negligible role in its precise determination.

\section{Summary and outlook}

We have presented next-to-leading order calculations of heavy-quark contributions to the unpolarized deep-inelastic scattering structure functions $F_4$ and $F_5$ within the ACOT scheme. Our study confirms that $F_4$ receives its first non-zero contribution at NLO, consistent with the AJ relations that predict a vanishing leading order term in the massless quark limit. In contrast, $F_5$ exhibits moderate but non-negligible higher-order effects, with deviations between LO and NLO predictions reaching up to about 10\%. 
These findings highlight the relevance of higher-order QCD corrections and heavy-quark dynamics, particularly at small values of $x$, where gluon and strange-quark contributions become significant. 

\begin{itemize}
    \item NLO corrections substantially modify the behavior of $F_4$ and $F_5$, emphasizing the importance of including heavy-quark effects in precision QCD analyses.
    \item The NLO calculation has been implemented in \texttt{APFEL++}, allowing for fast and robust numerical evaluations with modern parton distribution functions (PDFs).
\end{itemize}

Future work will extend this study by performing detailed comparisons between different heavy-quark schemes and by confronting theoretical predictions with upcoming and existing experimental data from facilities such as SHiP, IceCube, and DUNE. 
Such comparisons will provide valuable insights into the interplay between perturbative QCD dynamics and heavy-flavor production in DIS, improving our understanding of parton distributions and weak interaction structure functions.\\

A detailed presentation of the underlying calculations, together with an extended phenomenological analysis, will be reported in two forthcoming publications: one dedicated to the unpolarized DIS structure functions $F_4$ and $F_5$ \citep{Spezzano:2025unpolarizedtodo}, and another focusing on the polarized case, covering $g_1$, $g_4$, $g_5$, $g_6$ and $g_7$ \citep{Spezzano:2025polarizedtodo}.

\section*{Acknowledgments}

The speaker thanks the organizers for the kind invitation and
his nCTEQ colleagues for insightful discussions.
The work of P.R.~was supported by the U.S.~Department of Energy under Grant \mbox{No.~DE-SC0010129,} 
and by the Office of Science, the Office of Nuclear Physics, within the framework of the Saturated Glue (SURGE) Topical Theory Collaboration.
P.R. thanks the Jefferson Lab for their hospitality. This material is based upon work supported by the U.S. Department of Energy, Office of Science, Office of Nuclear Physics under contract DE-AC05-06OR23177.

\bibliographystyle{JHEP}
\bibliography{polACOT_references}

@article{Hou_2021,
   title={New CTEQ global analysis of quantum chromodynamics with high-precision data from the LHC},
   volume={103},
   ISSN={2470-0029},
   url={http://dx.doi.org/10.1103/PhysRevD.103.014013},
   DOI={10.1103/physrevd.103.014013},
   number={1},
   journal={Physical Review D},
   publisher={American Physical Society (APS)},
   author={Hou, Tie-Jiun and Gao, Jun and Hobbs, T. J. and Xie, Keping and Dulat, Sayipjamal and Guzzi, Marco and Huston, Joey and Nadolsky, Pavel and Pumplin, Jon and Schmidt, Carl and Sitiwaldi, Ibrahim and Stump, Daniel and Yuan, C.-P.},
   year={2021},
   month=jan }

@unpublished{Spezzano:2025unpolarizedtodo,
    author = "Spezzano, E. and others",
    title = "{Heavy Quark Contributions to the DIS Structure Functions $F_4$ and $F_5$ in the ACOT scheme at NLO}",
    url = "\url{https://indico.cern.ch/event/1436959/contributions/6352265/}",
    note = "in preparation, see \url{https://indico.cern.ch/event/1436959/contributions/6352265/}"
}

@unpublished{Spezzano:2025polarizedtodo,
    author = "Spezzano, E. and others",
    title = "{Heavy Quark Contributions to the DIS Structure Functions $g_1$, $g_4$, $g_6$, $g_7$ in the ACOT scheme at NLO}",
    url = "\url{https://indico.cern.ch/event/1436959/contributions/6352265/}",
    note = "in preparation, see \url{https://indico.cern.ch/event/1436959/contributions/6352265/}"
}

@article{IceCube:2024nhk,
    author = "Abbasi, R. and others",
    collaboration = "IceCube",
    title = "{Observation of Seven Astrophysical Tau Neutrino Candidates with IceCube}",
    eprint = "2403.02516",
    archivePrefix = "arXiv",
    primaryClass = "astro-ph.HE",
    doi = "10.1103/PhysRevLett.132.151001",
    journal = "Phys. Rev. Lett.",
    volume = "132",
    number = "15",
    pages = "151001",
    year = "2024"
}

@article{Collins:1998rz,
    author = "Collins, John C.",
    title = "{Hard scattering factorization with heavy quarks: A General treatment}",
    eprint = "hep-ph/9806259",
    archivePrefix = "arXiv",
    reportNumber = "PSU-TH-198",
    doi = "10.1103/PhysRevD.58.094002",
    journal = "Phys. Rev. D",
    volume = "58",
    pages = "094002",
    year = "1998"
}

@article{Aivazis:1993kh,
    author = "Aivazis, M. A. G. and Olness, Frederick I. and Tung, Wu-Ki",
    title = "{Leptoproduction of heavy quarks. 1. General formalism and kinematics of charged current and neutral current production processes}",
    eprint = "hep-ph/9312318",
    archivePrefix = "arXiv",
    reportNumber = "MSUHEP-93-15, SMU-HEP-93-16",
    doi = "10.1103/PhysRevD.50.3085",
    journal = "Phys. Rev. D",
    volume = "50",
    pages = "3085--3101",
    year = "1994"
}

@article{Aivazis:1993pi,
    author = "Aivazis, M. A. G. and Collins, John C. and Olness, Fredrick I. and Tung, Wu-Ki",
    title = "{Leptoproduction of heavy quarks. 2. A Unified QCD formulation of charged and neutral current processes from fixed target to collider energies}",
    eprint = "hep-ph/9312319",
    archivePrefix = "arXiv",
    reportNumber = "SMU-HEP-93-17, MSUHEP-93-17, PSU-TH-138",
    doi = "10.1103/PhysRevD.50.3102",
    journal = "Phys. Rev. D",
    volume = "50",
    pages = "3102--3118",
    year = "1994"
}

@article{Kretzer:1998ju,
    author = "Kretzer, S. and Schienbein, I.",
    title = "{Heavy quark initiated contributions to deep inelastic structure functions}",
    eprint = "hep-ph/9805233",
    archivePrefix = "arXiv",
    reportNumber = "DO-TH-98-05",
    doi = "10.1103/PhysRevD.58.094035",
    journal = "Phys. Rev. D",
    volume = "58",
    pages = "094035",
    year = "1998"
}

@article{Stavreva:2012bs,
    author = "Stavreva, T. and Olness, F. I. and Schienbein, I. and Jezo, T. and Kusina, A. and Kovarik, K. and Yu, J. Y.",
    title = "{Heavy Quark Production in the ACOT Scheme at NNLO and N3LO}",
    eprint = "1203.0282",
    archivePrefix = "arXiv",
    primaryClass = "hep-ph",
    reportNumber = "LPSC-12-048, SMU-HEP-12-05, KA-TP-09-2012",
    doi = "10.1103/PhysRevD.85.114014",
    journal = "Phys. Rev. D",
    volume = "85",
    pages = "114014",
    year = "2012"
}

@misc{shipcollaboration2022shipexperimentproposedcern,
      title={The SHiP experiment at the proposed CERN SPS Beam Dump Facility}, 
      author={{SHIP Collaboration}},
      year={2022},
      eprint={2112.01487},
      archivePrefix={arXiv},
      primaryClass={physics.ins-det},
      url={https://arxiv.org/abs/2112.01487}, 
}

@article{ParticleDataGroup:2024cfk,
    author = "Navas, S. and others",
    collaboration = "Particle Data Group",
    title = "{Review of particle physics}",
    doi = "10.1103/PhysRevD.110.030001",
    journal = "Phys. Rev. D",
    volume = "110",
    number = "3",
    pages = "030001",
    year = "2024"
}

@article{ALBRIGHT1975467,
title = {Neutrino production of M+ and E+ heavy leptons (I)},
journal = {Nuclear Physics B},
volume = {84},
number = {2},
pages = {467-492},
year = {1975},
issn = {0550-3213},
doi = {https://doi.org/10.1016/0550-3213(75)90318-1},
url = {https://www.sciencedirect.com/science/article/pii/0550321375903181},
author = {C.H. Albright and C. Jarlskog}
}

@article{Bertone:2016lga,
    author = "Bertone, Valerio and Carrazza, Stefano and Hartland, Nathan P.",
    title = "{APFELgrid: a high performance tool for parton density determinations}",
    eprint = "1605.02070",
    archivePrefix = "arXiv",
    primaryClass = "hep-ph",
    reportNumber = "CERN-TH-2016-103",
    doi = "10.1016/j.cpc.2016.10.006",
    journal = "Comput. Phys. Commun.",
    volume = "212",
    pages = "205--209",
    year = "2017"
}

@misc{abi2020deepundergroundneutrinoexperiment,
      title={{Deep Underground Neutrino Experiment (DUNE), Far Detector Technical Design Report, Volume II: DUNE Physics}}, 
      author={{B. Abi et al.}},
      year={2020},
      eprint={2002.03005},
      archivePrefix={arXiv},
      primaryClass={hep-ex},
      url={https://arxiv.org/abs/2002.03005}, 
}

@misc{khalek2022snowmass2021whitepaper,
      title={{Snowmass 2021 White Paper: Electron Ion Collider for High Energy Physics}}, 
      author={{R. Abdul Khalek et al.}},
      year={2022},
      eprint={2203.13199},
      archivePrefix={arXiv},
      primaryClass={hep-ph},
      url={https://arxiv.org/abs/2203.13199}, 
}

\end{document}